\documentclass[runningheads,a4paper]{llncs}

\usepackage{amssymb}
\usepackage{graphicx}

\usepackage{url}
\urldef{\mailsa}\path|trantrunghieu7492@gmail.com,  maxim.shcherbakov@vstu.ru|    
\newcommand{\keywords}[1]{\par\addvspace\baselineskip
	\noindent\keywordname\enspace\ignorespaces#1}

\usepackage{subfigure}
\begin{document}

\mainmatter  

\title{Detection and Prediction of Users Attitude Based on Real-time and Batch Sentiment Analysis of Facebook Comments}
\titlerunning{Detection and Prediction of Users Attitude}
\author{Hieu Tran
	\and Maxim Shcherbakov}
\authorrunning{H. Tran and M. Shcherbakov}

\institute{Volgograd State Technical Univeristy, Computer Aided Design Department,\\
	Lenin avenue 28, 400005 Volgograd, Russia\\
	\mailsa\\
	\url{http://www.vstu.ru/en}}

%


\maketitle

\begin{abstract}
The most of the people have their account on social networks (e.g. Facebook, Vkontakte) where they express their attitude to different situations and events. Facebook provides only the positive mark as a like button and share. However, it is important to know the position of a certain user on posts even though the opinion is negative. Positive, negative and neutral attitude can be extracted from the comments of users. Overall information about positive, negative and neutral opinion can bring understanding how people react in a position. Moreover, it is important to know how attitude is changing during the time period. 	
The contribution of the paper is a new method based on sentiment text analysis for detection and prediction negative and positive patterns for Facebook comments which combines (i) real-time sentiment text analysis for pattern discovery and (ii) batch data processing for creating opinion forecasting algorithm. 
To perform forecast we propose two-steps algorithm where: (i) patterns are clustered using unsupervised clustering techniques and (ii) trend prediction is performed based on finding the nearest pattern from the certain cluster. 
Case studies show the efficiency and accuracy (Avg. MAE = 0.008) of the proposed method and its practical applicability. Also, we discovered three types of users attitude patterns and described them.
\end{abstract}

\section{Introduction}

Sentiment analysis of textual content is used for opinion mining of people who express their emotions and thoughts by text messages. 
New communication platforms such as social networks (e.g. Facebook or VKontakte) gives a new opportunity for better understanding information using natural language processing and sentiment analysis. 
According to the article published by zephoria.com in December 2015, nowadays Facebook  has more than 1.55 billion monthly active users \cite{zephoria}. These users write more 510 000 comments every minute and this is a source of large information on the Internet. 
Usually, these textual comments are the results of the reaction of people regarding recent news or happened events. 
Understanding of users attitude helps to know how a certain person or groups respond to the particular topic, and it serves to draw relevant conclusions or make efficient decisions based on feedback \cite{LSSA,omasa}.
For example, in the political field. Assume there is news regarding a particular decision of the government in a certain country, which published in social networks by BBC or CNN. Based on examination of the textual comments, we can understand people positions and either a certain person supports this decision of the government or not. 

From the business point of view, sentiment analysis helps companies to improve customer development process, enhance business intelligence systems, and change their marketing strategies to get more profit. 
Moreover, using this type of text analysis, the trend of people’s attitude to certain events or typical groups of events can be predicted. This foresight is valuable for proactive actions development for the future expected situation in every domain we refer to, such as politic, economic, business and so on.
So, in fact, the questions is how to understand users behaviour and opinions according to processing textual comments in social network and how to predict either these opinions remain the same or will be changed in time?  
Opinions give the intuition about a person or customer preferences. 


The main problem, which is considered in the current research, is how to understand positive or negative user’s opinion about published posts and news using sentimental text analysis. Are there any laws and consistent timewise patterns in user’s comments, and how to detect these patterns and predict them?
The contribution of the paper is a new method based on sentiment text analysis for detection and prediction negative and positive patterns in Facebook comments.

The paper contains the following sections besides of introduction.  The next section contains the literature review and analysis of the recent related works on sentiment text analysis. After it describes the main idea of the proposed method of Facebook comments sentiment analysis using a combination of the real-time and batch data processing.  Results and discussion are covered in the last section.

\section{Related works}
Sentiment text analysis is a large but still growing research domain. An early, and still common, approach to sentiment analysis has been to use the called ‘semantic orientation’ (SO) of terms as the basis for classification \cite{PSO}. 

Turney showed that semantic orientation is useful for classifying more general product reviews \cite{tuotu}. The work suggests that product reviews may be classified easier than movie reviews, as product reviews tend to be shorter and be more explicitly evaluative. 
In \cite{tusc} authors classified movie reviews used standard bag-of-words techniques with limited success.
Twitter is a social network which represented as a sources of customer opinions to analyze. 
The early results of Twitter data sentiment analysis presented in work 
~\cite{csim}. The authors of the paper "Sentiment Analysis on Twitter" tentatively conclude that sentiment analysis for Twitter data is not that different from sentiment analysis for other genres~\cite{saotd}. 
In~\cite{wit} authors used the similar way (Twitter API) to collect training data and then to perform a sentiment investigation. An analysis of Twitter followers reveals networks of users who are related by current news tops rather than by personal interactions. Furthermore, a database of sensor data from the reality mining corpus is used for dynamic social network analysis~\cite{rmmc}.
Besides social networks, common websites have a large data needed to investigated. As practical implementation, sentiment analysis was applied to the feedback data from Global Support Services survey ~\cite{scocfd}. It helps organizations determine the quality of services. 
Text information from social networks provides information about entities (people and organizations) and their corresponding relations and involvement in events ~\cite{mndt}. 
Up to this time, there are not too many articles about real-time text data analysis. In the research~\cite{dsaf} proposed the theory of big data stream analytics for language modeling for sentiment analysis but did not put into practice with a certain network. 
A group authors from the University of Southern California described a system for real-time analysis of public sentiment toward presidential candidates in the 2012 U.S. election as expressed on Twitter ~\cite{asfr} but this system only works with the twitters, and could not reach to users replies. 

Almost all of research is concerned about status, twitter, or post to analysis sentiment ~\cite{cacv,apgm,tpta}.  There is a small number of research papers mentioned the replies and comments analysis. However, replies to Twitter and Facebook comments have values of sentiment expressions from users. 
In ~\cite{thai}, authors implemented analysis of Facebook posts ~\cite{wias} on the 2014 Thai general election. They did not use a large information from the comments of each post. Therefore, we focus on these comments which will be performed in this paper.
Also in ~\cite{asaos} presented prediction the vote percentage that individual candidate will receive on Singapore Presidential Election 2011 using Twitter data with census correction but they met issues with fake tweeter sentiment. And the source may be related to the scenario where the voters do not truly reflect their on-line sentiments from their choice of candidate.  Several case studies have found that the online information has been quite successful acting as an indicator for electoral success~\cite{wias}. There are some research on the use of twitter such as ~\cite{w140}.
There is the Vietnamese research group do analysis based on Facebook comments, but they worked in a sphere of Vietnamese language ~\cite{lbao}.

Based on literature review we conclude, that sentiment analysis of text comments on the social network as a mechanism to understand the people's attitude is still an open question.

\section{A method}
\subsection{General scheme}

The main problem we would like to solve is the creation of the technique, which helps understand and predict user attitude expressed in Facebook comments regarding published news or post. We propose the following method: (\textit{1}) using collected from Facebook data we perform batch analysis to make the precise forecasting model to detect and predict negative and positive behaviour patterns; (\textit{2}) using real-time text processing we detect a pattern and understand what is the current situation looks like; (\textit{3}) understand how this situation will be developed based on pattern prediction; and (\textit{4}) perform actions to change the pattern if it needed.

To be more concrete, we implement two main approaches for analysis sentiment in our method.  The first approach is real-time comments analysis. The solution allows to analyze and update results right in time when data generated by users on Facebook. The second approach analyses stored data in batch mode. This is a cluster-based approach for the negative/positive attitude patterns detection and prediction. To evaluate the performance of forecasting, the median absolute deviation measure is used.

\subsection{Real-time stream sentiment analysis and event generation}

\begin{figure}[t!]
	\centering
	\includegraphics[height=7cm]{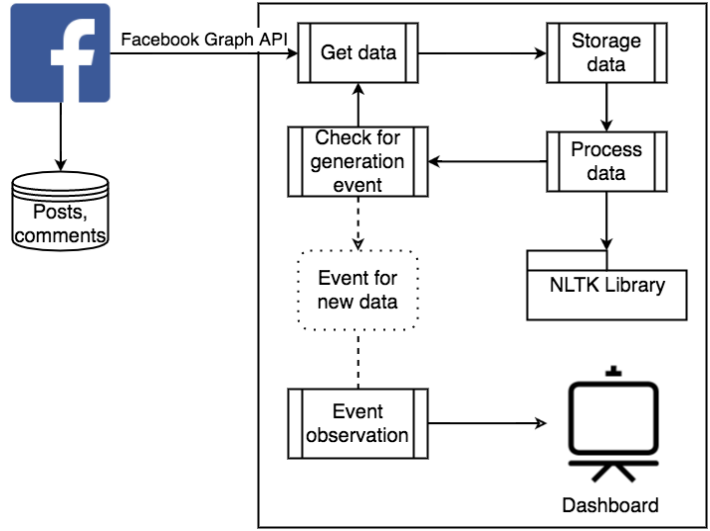}
	\caption{A scheme for real-time stream processing}
	\label{fig:1}
\end{figure}

Real-time stream processing solution retrieves data from Facebook server continually and then, processes a data package in minor period of time– almost real-time processing. 
The NLTK library is used for sentiment text analysis~\cite{nltk}. The results of data processing are checked by predefined user’s conditions. If it satisfies conditions, the solution creates an event to update dashboard’s status. 
Moreover, the real-time solution includes a procedure for listing to events. If a certain event occurred, the dashboard will be updated. Figure \ref{fig:1} presents a proposed scheme for real-time stream processing.

\subsubsection{Data retrieving.}
To keep a set of comments updated, a loop statement is used to receive comment data from Facebook. Time ($T_u$) between two loops statements is defined in advance. We set  $T_u$ = 0.1 seconds as it is nearly real-time and allows to store data continually updated. However, this parameter might be changed if needed.
Graph API Facebook is used in every step of the loop to obtain all comments of the selected post and to transfer data for storing and further analysis. This approach provides assurance, that collected data is fresh. 


\subsubsection{Event generation.}
In the inner loop, when data is transferred for further processing, the comparison of a new segment of data, which has been obtained from Facebook recently with cached data, is made. 
It is crucial as it helps us find out the changes in data. If changes in data are discovered, the event needs to be generated. Observers will receive this event further.

%

\subsubsection{Event Observation.}

To evaluate the performance of the method, the programm implementation includes the procedure which allows observing any generated event. In occasion of the event reflecting changing data, the solution updates system automates re-analyze sentiment data. This involves the change of dashboard status. Furthermore, cache data is updated as well to be sure we have newest data from the server. 
For instance, we analyse the sentiment of comments on a post about a political topic entitled 'Obama bans solitary confinement for juveniles and low-level offenders' on CNN news channel on Facebook\footnote{The title of the post: Obama bans solitary confinement for juveniles and low-level offenders, \url{https://www.facebook.com/bbcnews/posts/10153348871732217}
	}. 

\begin{figure}[t!]
	\centering
	\begin{subfigure}[]
		\centering
		\includegraphics[height=1.75in]{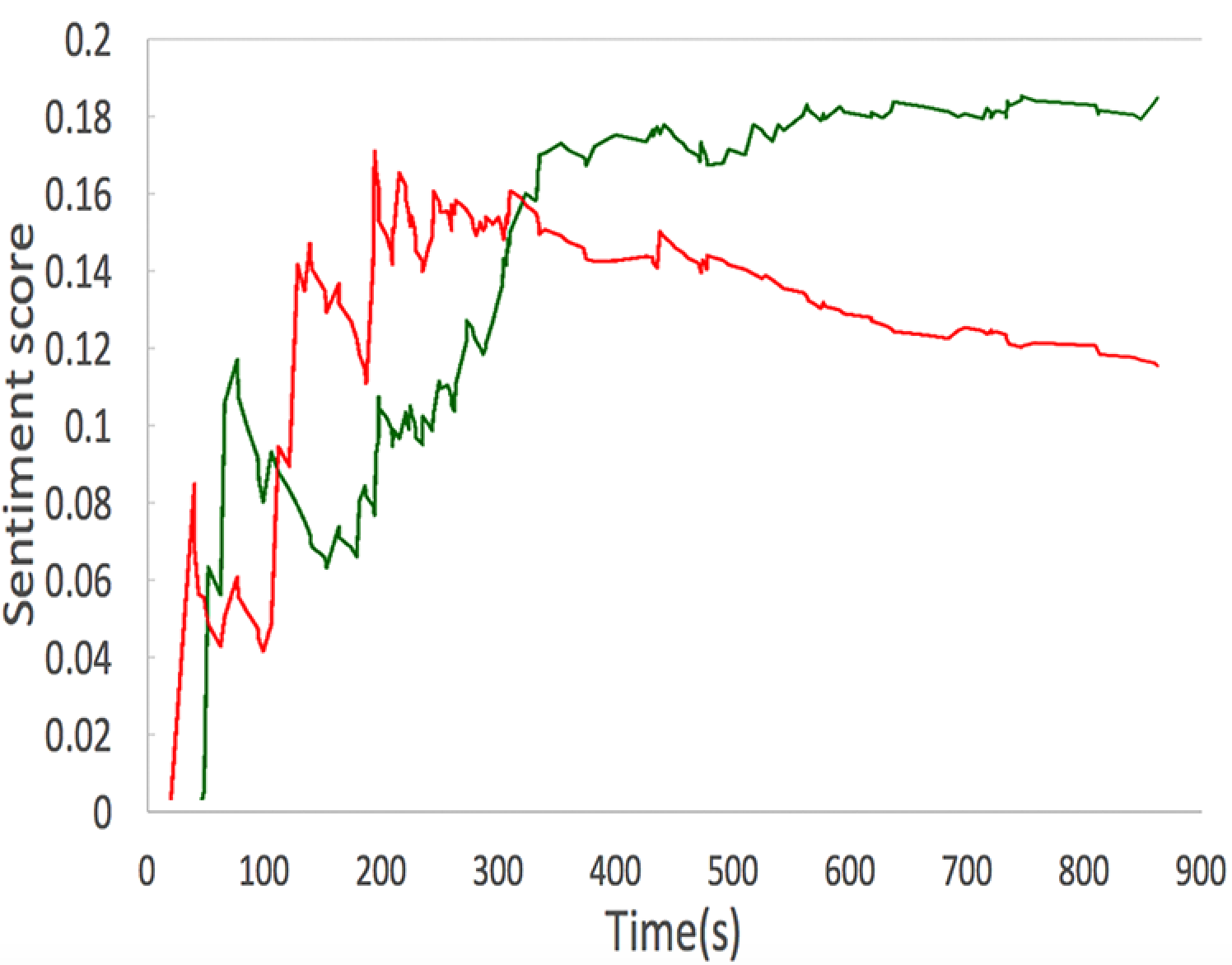}
	\end{subfigure}%
	\begin{subfigure}[]
		\centering
		\includegraphics[height=1.75in]{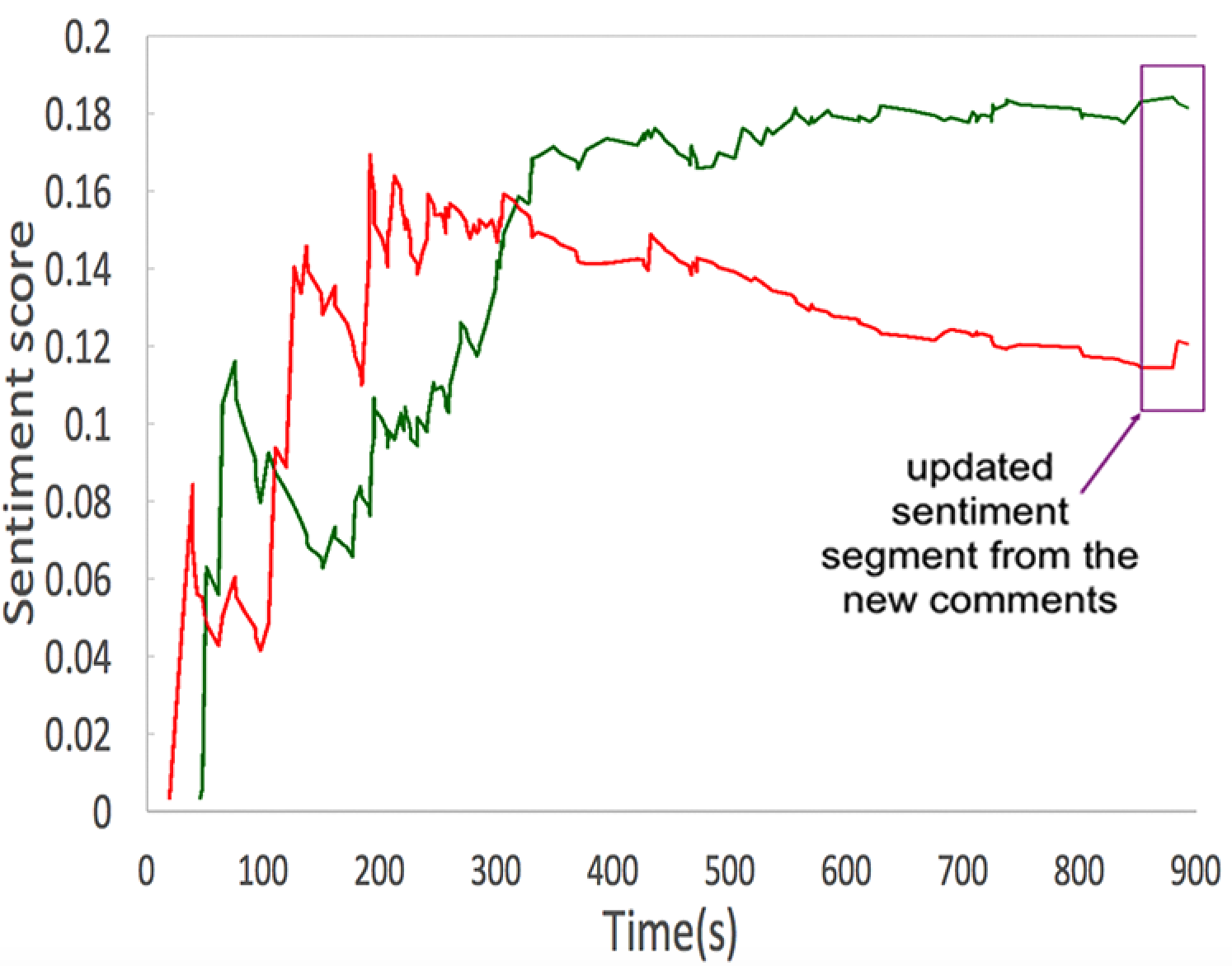}
	\end{subfigure}
	\caption{Dashboard represents (a) negative/positive sentiment of the first 750 seconds of post's life; (b) updated results after synthetic negative comment}
	\label{fig:4}
\end{figure}

Figure \ref{fig:4} shows the dashboard representing the negative and positive attitude in different timestamps. 
The red line represents values of negative sentiment. The green line is the positive’s. These values are fall into the interval $ [0; 1] $.  Every moment $ t $, we have a sum of negative, positive and neutral equals to 1:
\begin{equation}
V^{(p)}_{t} + V^{(n)}_{t} + V^{(u)}_{t} = 1,	
\end{equation}
where, 
$V^{(p)}$  - denotes the positive score, $V^{(n)}$  - denotes the negative score,
$V^{(u)}$  - denotes the neutral score.
In this paper, we do not place the neutral scores in the graph. The scale 'time' represents the time of comment posting as the interval from creating a post in seconds. The 'sentiment analysis' scale is the polarity value of comments.

The first figure explains the behavior for every 780 seconds of post life and the second one, reaction on the posted synthetic negative comment.
Real-time analysis allows to detect current patterns and to compare obtained pattern with expected or required. 
However, finding and adjusting those references patterns depends on expert (human) intervention and due to high velocity and the variety of data, this procedure is very costly.


\subsection{Batch data processing}

\begin{figure}[h]
	\centering
	\includegraphics[height=4cm]{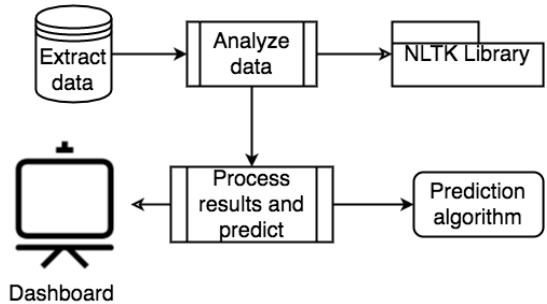}
	\caption{Scheme for batch processing}
	\label{fig:6}
\end{figure}

As method includes pattern detection and forecasting models, another component of our system implements pattern discovery in batch mode via processing of high volumes of certain topic data collected over a long period of time. Comments in the form of textual content regarding the topic are collected from several popular pages. The number of posts is selected arbitrary (100 posts, 1000 posts, or even more).  Our solution uses NLTK and the results are used for sentiment pattern detection and prediction. Figure \ref{fig:6} shows the scheme for batch processing of sentiment analysis.

\subsubsection{Data collecting}

Implementation of batch data processing makes sense in the case of high volumes data. 
Firstly, we chose a topic, which is popular recently. 
For each post, using Facebook Graph API, all comments have been collected during the first 30000 seconds. Data is stored in flat table format (e.g. CSV file) which is easy to save in distributed file system. 
The header of CSV file contains the following columns: 
[Datetime] [Topic] [Post] [Comment] [Positive] [Negative].

%
%

The first column contains a value in seconds when a comment was appeared, which is counted from initial post’s appearance. Data type of the value is the integer. 
The second column contains the topic’s title. The third column is post’s title. The fourth column is the content the text of a comment. The last two columns represent sentiment analysis scores (negative and positive) of comment on the post. The data type of scores is the float number. 

\subsubsection{Analysis data.}

\begin{figure}[h]
	\centering
	\includegraphics[height=5cm]{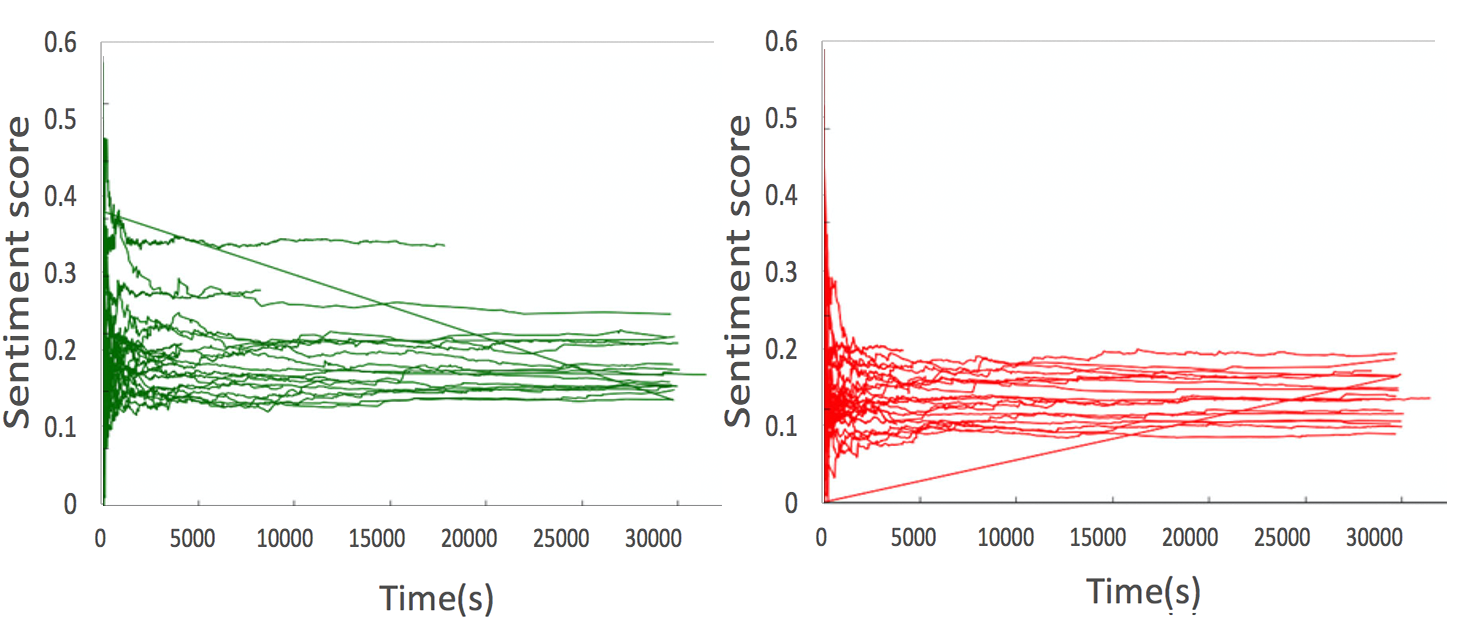}
	\caption{Positive sentiment  time series (left side) for a set of posts and negative sentiment time series (right side). Data about U.S. presidential election 2016 topics was obtained by CNN and BBC feeds}
	\label{fig:9_1}
\end{figure}

As it was mentioned above, the NLTK sentiment analysis was applied for each comment for the post. As the results, we obtain time series which describes people’s negative $V^{(n)}$ and positive  $V^{(p)}$  scores during the time. Figure \ref{fig:9_1} expresses positive scores time series (left side) for a set of posts and the negative (right side).

%

\subsubsection{Clustering data.}

The output of the previous step is a set of time series and the number of time series is equal to a number of examined posts. 
We use the fact, that some of the time series has a similar pattern and they could be arranged into 3 or 4 different groups. It allows defining a 'typical' behaviour of the people from a sentiment point of view. For instance, the certain post could have a high negative expression in the beginning and fade negative afterwards. 
The well-known technique for unsupervised grouping is clustering.
In this paper, two machine learning approaches: $ k $-means and MB-means were used to cluster sentiment of posts. 
Needless to say, $ k $-means has been studied and applied in a wide range of domains, e.g. transportation network analysis~\cite{strategway}, information security~\cite{aon}, pattern recognition~\cite{fsf}, text classification~\cite{tusc} and many others domain.

\begin{figure}[t!]
	\centering
	\includegraphics[height=1.40in]{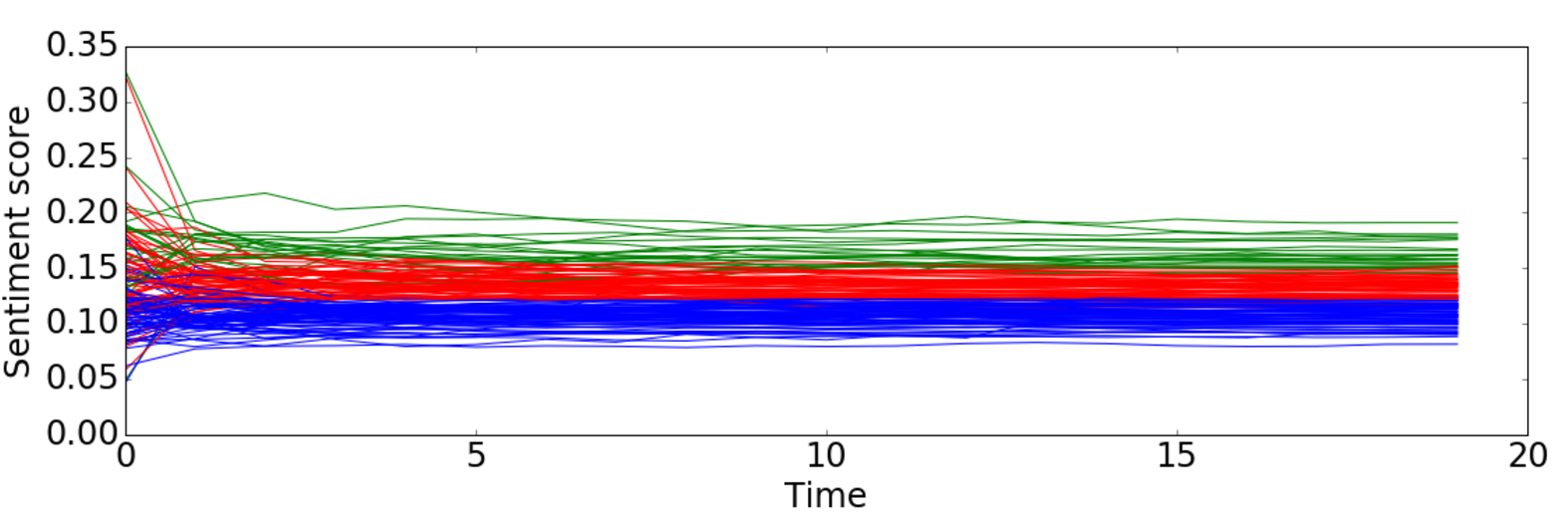}
	\caption{The results of clustering using (a) $ k $-means algorithm and (b) MB-means algorithm, where every plot has color according to the cluster}
	\label{fig:11}
\end{figure}

We implement clustering, where initial time series are placed in three difference groups according to their characteristics. To perform clustering, each of posts expressed as a vector with 20 components (features): [id][Post][Period 1][Period 2]... [Period 20].

These features describe to our first 30,000 seconds of the posts. The values of  30,000 seconds  have been chosen arbitrary and might be changed if needed. Every feature has the same period time, which is about 1,500 seconds. The value of each feature is the average value of all value in this period. 
All-time series (or vectors) have been divided into three clusters using $ k $-means algorithm. The number of clusters picked up according to preliminary analysis.
For comparison with $ k $-means method, we also use the other algorithm to cluster which is named MB-means. 
Figure \ref{fig:11} shows the results of clustering, every time series has the color according to the cluster.
However, in spite of difference clustering techniques, the results look quite similar. 

We can describe three clusters in the following way.
\begin{itemize}
\item The first cluster which is red lines (based on \textit{k}-means algorithm results) begins with high positive scores. It decreased quickly from the beginning to the third period, then it continuously went downward slowly. From the fifth period, it felt the lowest value. Then it remained stable in the next periods.   
\item With the second cluster which is green lines using $ k $-means, we have a graph remained relatively stable from the beginning. Almost periods time, there were small increases or decreases. 
\item The last cluster is blue lines using $ k $-means. It started with low positive scores. It grew up rapidly, and then it reached the peak of the score at the second period. Then there were slightly drop in the third period and leveled off. From the next period, it stayed constant and there are no more changes.
\end{itemize}

In conclusion, each cluster has characters itself obviously. It is the good signification to cluster testing data. 

%
%
%

\subsubsection{Prediction.}

Clustering allows defining typical patterns in people's behaviour.  The next task is prediction the trend development of people’s attitude on a post using data about the first reaction (or 5 values in vector representation).  Note, we choice 5 values for the current research, but this parameter is subject to choice. 

The prediction is performed in two steps. 
The first step including cluster detection procedure. Based on the first 5 features (or 5 values in vector representation) of given pattern which needs to be predicted, we select the appropriate cluster. The selected cluster is the cluster having the average profile for this 5 features with minimum deviation with our post (in terms of Euclidean distance).
In the second step, we predict the trend development of people’s attitude on this post. The nearest (in terms of distance) time series from the certain cluster is found and use as the prediction for the rest time interval.
The Mean Absolute Error (MAE) is used to evaluate the performance of the forecasting technique. 
\begin{equation}
	MAE = \frac{1}{n_{test}}\sum_{i=1}^{n_{test}} \left(  \frac{\sum_{j=6}^{20} |h_j - h_j^*| }{15} \right)_i,
\end{equation}
where, $h_j$  -- denotes the real value at the $ j $-th timestamp, $h^{\ast}_j$  -- denotes the predicted value at the $ j $-th timestamp, $ n_{test} $ -- a number of time series included in test data set.

%




\section{Results and discussion}

To evaluate our approach we designed and implemented software solution using Python and Facebook API. For our experiments, we used the topic “the United States presidential election 2016”. 
All data was received from two famous new channels: BBC news and CNN on Facebook. We collected 200 posts about the mentioned above topic. 
For every post, using Facebook Graph API, all comments have been collected during the first 30,000 seconds. Approximately, the total number of comments is about 100,000 comments. 

The results of real-time sentiment analysis indicate user’s emotions and thoughts in a real-time stream. Based on monitoring of dashboards, we can understand that the ratio of positive sentiment is higher that the negative sentiment. That means that the proportion of people who supports the news is more than the proportion who against. 
At the time when the intersection of the red line and the green line is observed, it’s the point that the trend of attitude is changed. It might the point that perhaps people are changing the attitude to the topic. Based on these points, we can create a line which is the general trend of people’s attitude. Also, it is possible to set up triggers indicates when an interested event occurs. 

\begin{figure}[t]
	\centering
	\includegraphics[height=4cm]{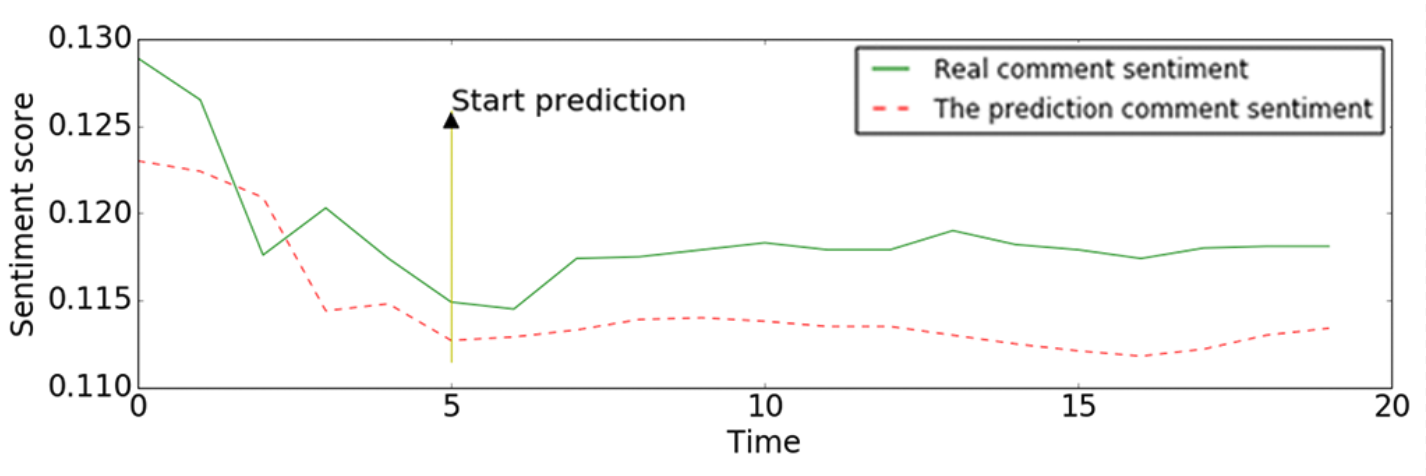}
	\caption{Prediction trending development of positive sentiment}
	\label{fig:14}
\end{figure}

Using clustering techniques we are able to detect the most typical behaviour of the users and describes them. For instance, we observe 
that negative or positive estimations asymptotically approaching to a certain level and never exceed the threshold. Negative and positive attitude is fading during the time, and we are able to estimate time of popularity of the post and advise actions to support popularity. 
Also, our technique allows predicting the trend development of people’s attitude. It could be a framework to detect the outliers in the comment of Facebook’s community. To evaluate forecasting performance MAE has been applied as error measurement. To avoid a case where results obtained by chance, we developed cross validation (with folds  = 10) and get average MAE =  0.008. 
Figure \ref{fig:14} gives the representation of results of trend forecasting. The green line is the real comment sentiment on this post. And the red line is the prediction line for the development of people’s attitude on the post.\\

%

%
%




\section{Conclusion}

In this study, we perform actions to understand users preferences and attitude based on sentiment analysis of Facebook comments and application of machine learning techniques. 
Detection of laws and consistent patterns in users comments published in time framework allows providers of services and sales to react in real time and be more proactive using trend prediction. 
We propose a new method based on sentiment text analysis for detection and prediction negative and positive patterns for Facebook comments which combines (i) real-time sentiment text analysis for pattern detection and (ii) batch data processing for creating forecasting algorithms. 
To perform forecast we propose two-steps algorithm where: (i) patterns are clustered using unsupervised clustering techniques and (ii) trend prediction is performed based on finding the nearest pattern from a certain cluster. 

%

Based on the results, we found three types of user behavior in their opinion expression and find that our simple forecasting technique is very accurate. Proposed method can be readily used in practice by sales companies who can use the real-time approach for learning their customer attitude about products and  making the assessment of a product. 
Some social and political organizations use to analyze community on a certain event such as The 2016 U.S. election.
Our future work will be continued by focusing on improvement the model training, improvement method of prediction using MAE. Besides, the next stage of our research will analyze a group of people such as Vietnamese and Russian \cite{novt,amats}.
\subsubsection*{Acknowledgments.} The reported study was partially supported by RFBR, research project No. 16-37-60066 and research project MD-6964.2016.9.


%
%

%
%
%
\end{document}